# VECTORS

**VECTORS – Video communication through opportunistic relays and scalable video coding**

**Abhishek Thakur, Arnav Dhamija, Tejeshwar Reddy G**
BITS Pilani, Hyderabad Campus, Jawahar Nagar, Hyderabad, Telangana, India 500078
abhishek@hyderabad.bits-pilani.ac.in; arnav.dhamija@gmail.com; tejeshwarreddy1996@gmail.com

**Abstract.**
*Crowd-sourced video distribution is frequently of interest in the local vicinity. In this paper, we propose a novel design to transfer such content over opportunistic networks with adaptive quality encoding to achieve reasonable delay bounds. The video segments are transmitted between source and destination in a delay tolerant manner using the Nearby Connections Android library. This implementation can be applied to multiple domains, including farm monitoring, wildlife, and environmental tracking, disaster response scenarios, etc. In this work, we present the design of an opportunistic contact based system, and we discuss basic results for the trial runs within our institute.*

**Keywords:**
*User-generated video content; opportunistic communication; user mobility; disruption tolerant network*

*1. Motivation and Significance*

In recent years, user-generated video content has increased several-fold [1]. In most scenarios, the content is captured from mobile devices and uploaded to servers for subsequent viewing. However, such solutions do not work in the absence of network infrastructure. Delay and Disruption Tolerant Networks (DTNs) [2] can be applied in such scenarios for communication. Opportunistic Networks (OppNets) are a special case of DTNs in which the contact patterns are unpredictable. Present approaches that try to send video at fixed quality, which may lead to high delay or drop rates. If the video flow is not adaptive, the network may not be optimally utilized.

Common Internet-based approaches for streaming video (e.g., DASH/TCP) rely on the end-to-end transmission for video quality adaptation. When the round-trip times are below a second, these approaches perform well. However, such adaptations will not suit OppNets primarily on two counts: 1) the feedback from the destination can take an undeterministically long time to reach the source; 2) some of the feedback about the acknowledged packets from the destination may be lost.

OppNets create multiple copies (replicas) of the data to reduce delivery delay and improve delivery probability. However, the creation of too many replicas can overload the network. Hence we need to devise a solution that delivers a reasonably good quality video without overloading the network. Additionally, when the OppNet has more available network resources (e.g. buffer space, number of nodes, frequency of contacts), the delivered video quality should improve accordingly. On the other hand, when network lacks resources, a lower quality video should still be delivered.

In this paper, we present "**V**id**E**o **C**ommunication **T**hrough **O**pportunistic **R**elays and **S**calable video coding" – a.k.a. VECTORS, a software for carrying scalable video over OppNets, deployed using commodity Android devices. Scalable Video Coding (SVC) [3,4] compresses the segments of video into payloads at different SVC layers. Section 2 discusses the SVC compression algorithm in more detail. The video source in VECTORS adapts the number of SVC layers being transmitted, to optimize the delivered video quality.

**1.1 Background and Prior Work:**

Table 1: Key terminology for video communication over OppNet.

| Term | Definition |
| --- | --- |
| Acknowledgment | This is a list of payload-identifiers that have been received by the destination. Nodes, delete local copies for such payloads. |
| Copy-count (L) | The number of copies (replica) that can be created for the payload. The source node sets initial value of L. |
| Destination | The node that will decode the video segments. |
| Node | A computer system, which may be mobile or static. |
| Opportunistic Contact | When two nodes are in vicinity of each other and capable of exchanging data. |
| Payload | The unit of application-level data being communicated through OppNet. |
| Relay | Transfer of a payload to a node that does not have it. This can create additional copies in OppNets. |
| Segment | The interval of video on which compression in done at source. |
| Source | The node that is capturing and compressing the video segments. |
| SVC-Layers | A video segment will generate multiple payloads for transmission. It includes base-layer and multiple enhancement layers. |
| TTL | Time-to-live – the duration for which the OppNet will store the payload (from creation time). |

SVC compresses the video into multiple layers. The base layer may have a lower frame rate, lower resolution, higher quantization error, and lower bits-per-sample. One or more enhancement layers improve temporal (higher frame rate), spatial (better resolution), quantization (lower loss) or bit-depth. Each SVC layer may be transmitted separately. The destination needs the base layer and contiguous enhancement layers for decoding to succeed. For example, if the base layer is not delivered, the particular segment cannot be decoded, even if all other enhancement layers are received.

OppNet routing uses the principle of Store-Carry-Forward to deliver the payload from source to destination. It typically uses multi-copy routing to improve the probability of delivery and to reduce the delivery delay. Epidemic routing [5], allows uncontrolled replication of payloads and it can provide the best results when the network is lightly loaded. As network load increases, creation of additional copies can lead to congestion and may negatively impact delivery. Multiple routing protocols have been proposed to control the overheads of creating several copies. Spray-And-Wait [6] (SNW) uses a simple approach wherein the source node controls the maximum number of copies that will exist on the network. In Binary SNW protocol, once a payload is relayed, both the nodes participating in the relay will reduce the copy-counts of the payload by half. When the copy count reaches one for a particular payload, the node can only deliver the payload to the destination.

In most OppNets, it is frequently the case that some message payloads may not reach the destination. Transferring standard H.264/H.265 compressed video using such networks may result in several segments of video being lost entirely. On the other hand, SVC compressed video is amenable for transferring over OppNets as SVC does not require all the layers of the video to be transmitted for decoding. As more nodes come into contact with the destination node, there is a greater chance for more SVC enhancement layers to be transferred and the decoded video quality to improve.

Lenas [7], Morgenroth [8] and Blanchet [9] have demonstrated video streaming over DTN for controlled network settings. These demonstrations rely on specialized nodes (e.g., satellites and static nodes with round-trip delay below 2 seconds) to help complete the video flow over DTN. Hence they do not work for OppNets where the contact patterns are unpredictable.

As done in [8], we experimented with IBR-DTN [10] for communication between Android devices. In the absence of access points, IBR-DTN can work by creating an ad-hoc Wi-Fi Direct [11] connection. The Android implementation for Wi-Fi Direct in real-world deployments is challenging, as it requires frequent human interactions for acquiring user permissions.

**1.2 Experimental Setup**:

A deployment of VECTORS consists of a source and a destination node as shown in Fig. 1. Multiple Android devices running the VECTORS application are used to create the opportunistic network.

The software at the source node records the raw video segments for five minutes before compressing it using SHM [12] to create multiple SVC layers. The source node extracts each SVC layer as a separate payload. The extraction information is also transmitted to the destination using VECTORS.

The software at the destination (the decoder) combines the received payloads using the extraction information. Note that the lower layers and the extraction information are necessary to decode higher SVC layers. The destination software also generates acknowledgment for all the payloads that it receives.

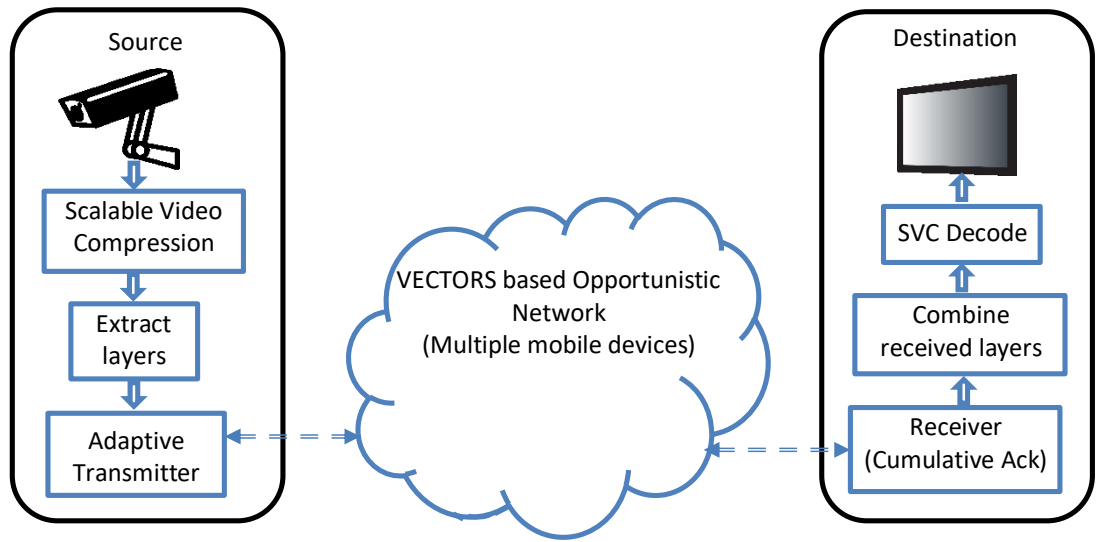

Figure 1. Data flow for SVC content using VECTORS

For our deployment, both source and destination are Linux computers since at the time of writing, SHM does not presently have an Android port. We connect an Android device to both the source and the destination. These computers use the Android Debug Bridge (ADB) to transfer the payloads to/from mobile phones over the USB interface.

## 2. *Software Description*

The software for VECTORS has three main components; 1) the adaptive compression, extraction, and transmission at the source; 2) the VECTORS Android application for the OppNet and 3) the combiner and decoder at the destination.

**2.1** *Software Architecture*

**2.1.1** *Source*

The source node uses a cronjob to run a script to capture the raw video. It subsequently encodes the captured video using SHM. The SVC layers are extracted using the "ExtractAddLS" executable. Based on the acknowledgments that have been received, the push_at_src.sh script adjusts the number of SVC layers to be sent. It checks for acknowledgments for segments transmitted in the previous 6 hours, 12 hours and 24 hours. If all the three segments are acknowledged it increases the number of SVC layers that can be sent. If none of them are acknowledged, it reduces the number of SVC layers. If some of the segments are acknowledged, it does an additive increase / multiplicative decrease based on the number of layers being transmitted. For each payload transmitted, the source node assigns the maximum number of replicas (L) that may exist on VECTORS nodes.

**2.1.2** *Destination*

The destination node downloads all the payloads from the attached mobile device using a cronjob. It also uploads an acknowledgment file including the timestamp for the newly received payloads.

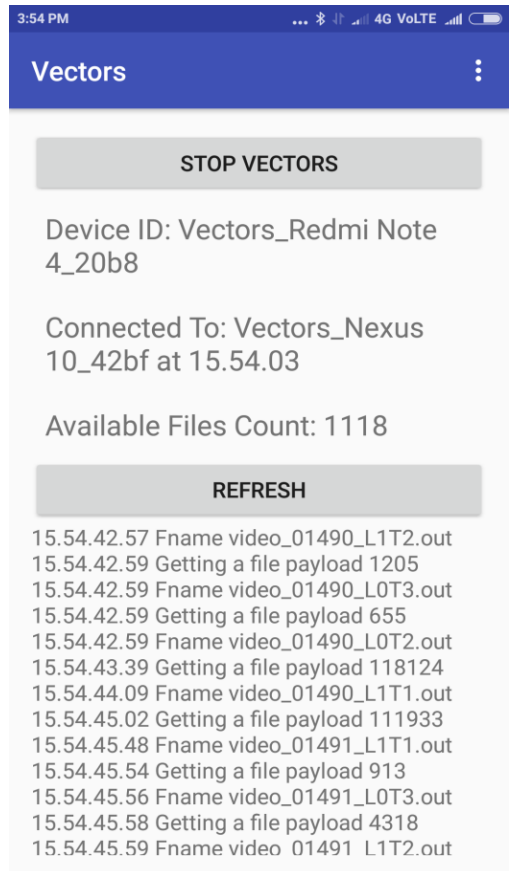

Figure 2. User Interface for the VECTORS application on Android nodes

**2.1.3** *OppNet*
The Android application implementing VECTORS OppNet has three main components:

- MainActivity - this is the main activity with a minimal user interface to enable/disable the VECTORS communication. Apart from this, it displays the communication state and logs as shown in Fig. 2.
- VECTORS Service - this is a background service implementing the relay of payloads between nodes using the state diagram illustrated in Fig. 3.
- StorageModule - this implements the storage and management of payload within the node. This module deletes expired payloads (TTL is expired).

Besides the above three classes, VECTORS Android app has additional classes to serialize/deserialize the metadata and ACKs. It also handles life-cycle events like restarting the VECTORS Service after the phone boots up.

**2.2** *OppNet State Diagram*

OppNet nodes, in this case, Android smartphones, are connected to each other using the VECTORS Android app. As the connections are opportunistic, the connection could be terminated any time due to the two nodes moving out of range (denoted by dotted lines in Fig. 3). Hence, the payload transfers can be considered to be best-effort. Our goal is to maximize the number of successful SVC layer transfers between two connected nodes over the opportunistic contact.

Nodes send video payloads only if the value of L is greater than one ("Transfer files" state in Fig 3). Acknowledgments are shared to all the nodes and are not constrained by copy counts.

**Discover and Connect**
Nodes broadcast their auto-generated unique ID over Nearby Connections [13]. On discovery of another node, it checks the ID of the remote node to decide whether a connection should be established. If the connection is successfully established, the node moves to Connection Initialized state. Otherwise, it remains in Discovery, attempting to connect to other nodes.

Occasionally, two nodes can have the same set of payloads because of prior contacts. To avoid such spurious connections in which no payloads are transferred, the connection is rejected for nodes with successful contacts in the last five minutes. Node mobility and other environmental conditions may cause the connection to fail, despite a node initiating the connection and the other node successfully accepting it.

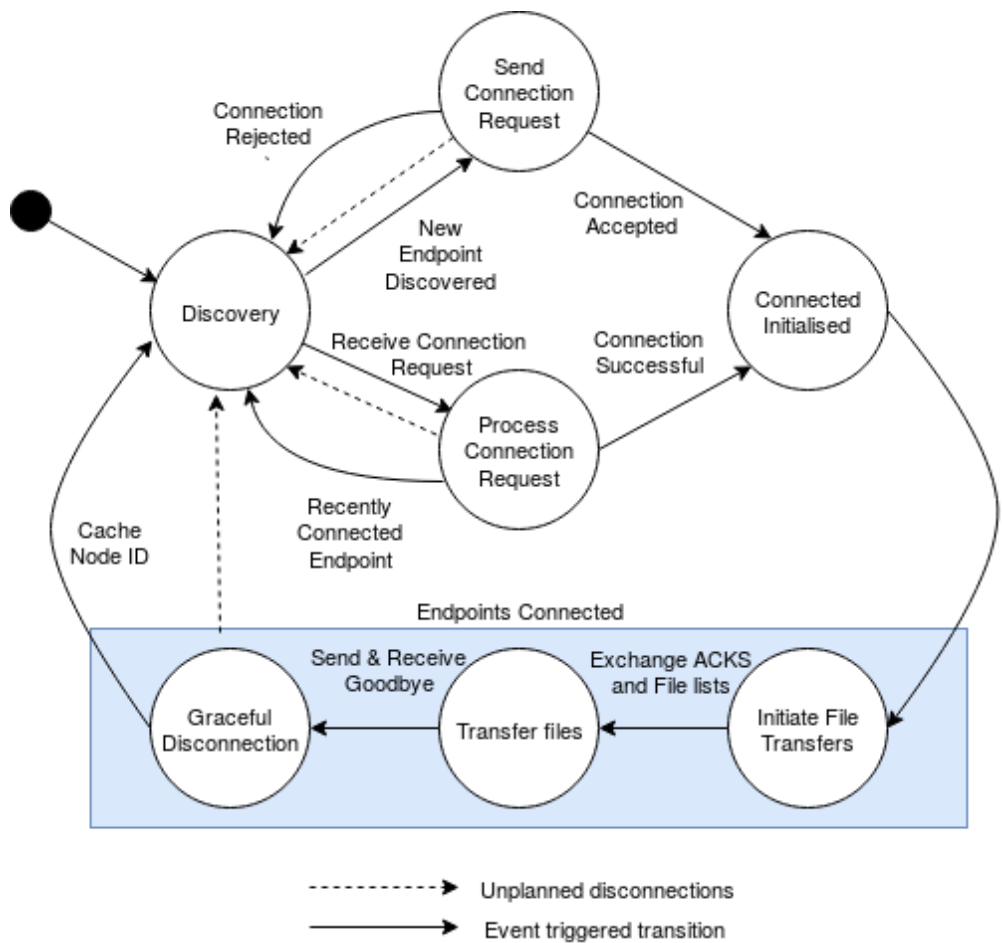

Figure 3. VECTORS Connection States (VECTORS Service)

**Endpoints Connected**

Once both the nodes connect to each other, they exchange a set of control messages, before transferring payloads. Each control message contains a four-byte header indicating the type of message followed by the message contents.

Firstly, they exchange the Acknowledgement (ACK) from the video destination. If the received ACK is newer than the current one on the node, it replaces the older ACK. Payloads acknowledged in the incoming ACK are deleted on the intermediate nodes.

In the next step, the nodes exchange a list of the payloads available with each other. From the list received, the node identifies the set of payloads to be requested. Based on the requested list, each payload is successively transferred with its VECTORS metadata. It transmits payloads in the descending order of their copy-count. The VECTORS metadata tracks the present copy-count of the payload and the nodes the payload has traversed. After successfully transferring the payloads, the metadata is updated and the copy-count is halved at both the nodes.

**Connection Termination**

Once a node has successfully received all the files, it sends a control message to inform the other node. A "graceful" disconnection occurs when both nodes exchange this control message.

As the network is ad-hoc and the nodes are mobile, there is a high chance of unplanned disconnections. For such unplanned disconnections, not all the payloads are transferred successfully. In this case, the nodes revert to the Discovery state without caching each other's ID in the recently connected list. This allows the pair of nodes additional attempts to exchange these payloads on reconnection.

3. *Illustrative Examples*

For initial experiments with VECTORS both source and destination were installed in a lab. Subsequently, we moved the source and destination to different corners of our academic building. In this stage, the devices of developers were used for testing the deployment akin to fire-fighters responding to an emergency situation in the corner of a building. After the issues related to the integration of encoding/decoding blocks and adaptation were resolved, the setup was moved to two different locations within our campus as shown in Fig. 4 below.

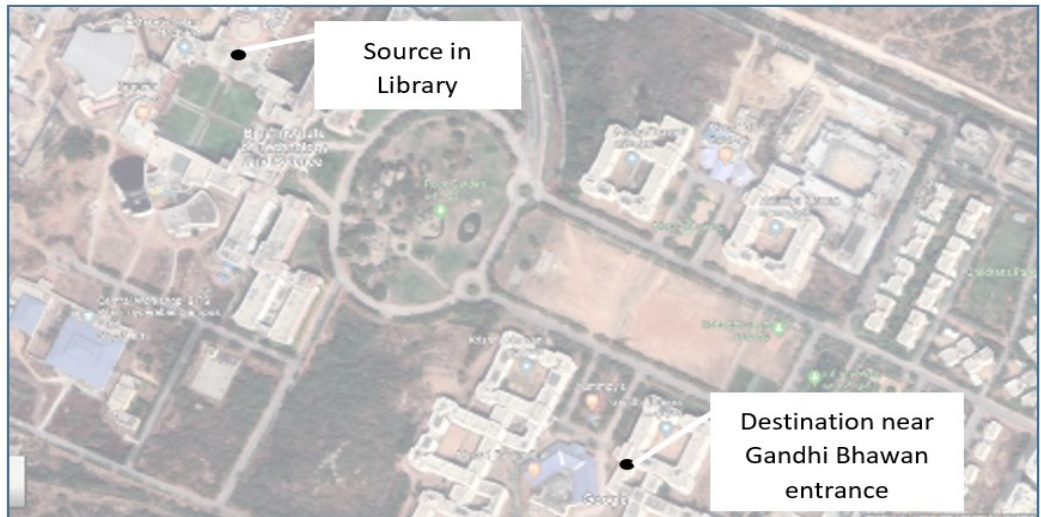

Fig. 4 Deployment within our Campus

The nodes are separated by a walking distance of 1.2 km. We invited the volunteers to install the VECTORS app on their devices. Over a two-week interval, 15 active nodes (each with more than 20 opportunistic contacts), helped complete the network.

## 4. Empirical Analysis

Since VECTORS operates in an open environment, it's not feasible to recreate identical experimental setup for different approaches to transmit video. To compare the performance of adaptive SVC transmission, we simulated different network load and node behaviours in the ONE simulator [14]. The simulation uses the contacts logged for the most active week of the deployment in Fig. 4. The captured video in raw format (YUV), was compressed at different video resolutions (Low - 320x240, Medium - 640x480 and High - 1280x960). For payload size, the simulation uses the file sizes for H265 (standard compression, without scalability) video and for SVC layers (using SHM). Fig. 5 shows the number of successfully delivered segments on Y axis, for various TTL values on X axis. For Fig. 5(A), we ran simulations for different video resolutions, with and without adaptive-SVC transmission. In Fig 5(B) we simulated higher disruption by removing nodes with maximum contacts, for Low-resolution video transmission. One, two and four nodes with highest contacts were dropped from the simulation. Adaptive SVC performs better than non-SVC as the nodes are removed, showing that it is more resilient to disruptions in the network. For brevity, further analysis of figures is left out of this paper.

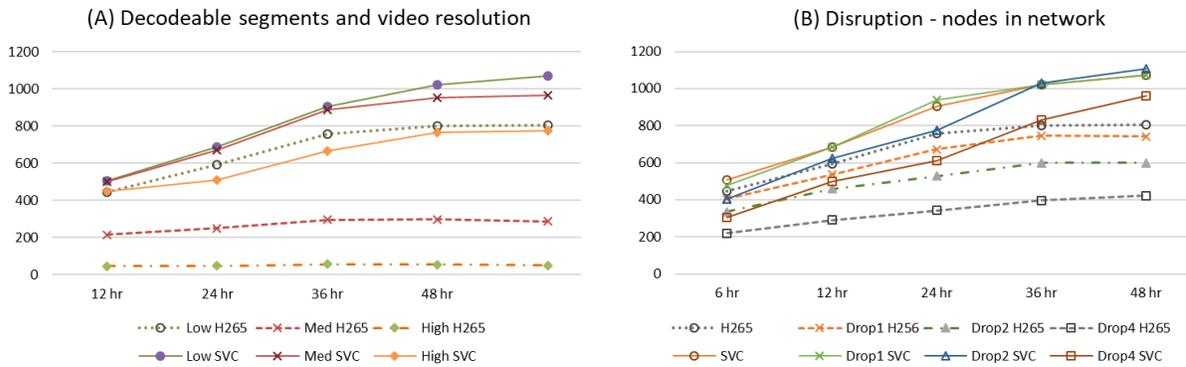

Fig. 5 Count of contacts between nodes

## 5. Impact

Depending on the mobility patterns of the devices and the distance between source and destination, the delays can vary from an order of minutes to order of hours. In a controlled setup, it is possible to transmit video with delays of the order of a few seconds. This can provide an affordable alternative to expensive network links (e.g. leasing satellite bandwidth) when real-time communication is not required. We discuss three sample scenarios below. Numerous other applications can be thought of, thanks to the increasing affordability of the Android devices used by VECTORS.

**Wildlife monitoring**: In the past, wildlife monitoring was done using cameras from which the data is manually extracted - a time-consuming and labor intensive approach. Prior research (Zebranet [15]) has utilized location and other contact details to be collected by using motes embedded in collars of animals. It may be possible to collect videos from across the areas where these animals roam. A solar powered, static video source can be deployed in the wild.

**Behavioral analysis for a transitory gathering of people**: Onnela [16] studies population dynamics for large transitory gatherings using messaging and call data records. Since many of these people are also expected to carry mobile devices, VECTORS can help provide a video feed for better contextual analysis of the events and people behavior. Another possible application in a similar realm relates to an improvement of law enforcement using VECTORS based communication.

**Rural deployments**: The developers intend to use VECTORS in agricultural environments to help track cattle in a field using cameras mounted on poles or trees with a solar-powered video capture device. Further, we aim to use it for monitoring the activities like the harvesting of fruits, flowers, and crops. The presence of such affordable infrastructure is also expected to discourage theft and reduce the need for the workforce to guard many of the farms and orchards.

## 6. *Limitations and future work*

The current implementation of VECTORS uses x86 Linux computers for the source and destination for processing video. The connection of Android devices to these computers via ADB complicates the deployment of VECTORS. Moreover, the SHM encoder is not parallelized and it takes a lot of time to encode high-resolution video. The present SVC adaptation only utilizes additive-increase / multiplicative-decrease. More advanced adaptation algorithms, including machine learning approaches, are open topics for research.

In future iterations of this project, we plan to port SHM to Android for a simpler setup of VECTORS. We also plan to improve the interoperability of VECTORS by implementing the Bundle protocol [17]. Lastly, we intend to experiment with alternate opportunistic routing protocols. Other possible extensions to VECTORS include the enforcing of a storage quota on each device and the addition of security in communication.

## 7. *Conclusions*

We were able to successfully have a media flow for the two-week interval with delays in playback varying between a couple of hours to a day. We had fifteen active participants, which is representative of rural deployments. The initial experiment shows that in the true rural scenario, VECTORS can be effective for communicating such content since these users will not have infrastructure based Wi-Fi.

Appendix A

The contact traces and simulation settings are publicly made available at https://github.com/swifiic/the-one.

**Required Metadata**

**Current code version**

| Nr | Code metadata description | Metadata |
|---|---|---|
| C1 | Current code version | v1.3.0 |
| C2 | Permanent link to code/repository used of this code version | https://github.com/swifiic/Vectors |
| C3 | Legal Code License | GPLv2 |
| C4 | Code versioning system used | git |
| C5 | Software code languages, tools, and services used | Java, Android, Linux Shell, C, C++ |
| C6 | Compilation requirements, operating environments & dependencies | Android 8.0 SDK, Google Play Services, Android Nearby Connections Library, Ubuntu 16.04 |
| C7 | Link to developer manual | https://github.com/swifiic/Vectors/wiki |
| C8 | Support email for questions | apps4rural@gmail.com |

**Current executable software version**

| Nr | Software metadata description | Metadata |
|---|---|---|
| S1 | Current software version | v1.3.0 |
| S2 | Permanent link to executables of this version | https://play.google.com/store/apps/details?id=in.swifiic.vectors |
| S3 | Legal Software License | GPLv2 |
| S4 | Computing platforms/Operating Systems | Android, Linux |
| S5 | Installation requirements & dependencies | Android 7.0 and above, SHM 12.4, JSVM, Ubuntu 16.04 |
| S6 | Link to user manual | https://github.com/swifiic/Vectors/wiki |
| S7 | Support email for questions | apps4rural@gmail.com |